
\magnification=\magstep1

\overfullrule=0pt
\def\mn{\medskip\noindent}

\def\w{$\cal W$}

\def\wh#1{\widehat{#1}}

\font\gro=cmr12 scaled\magstep2
\font\mit=cmr8 scaled\magstep2

\def\q#1{\lbrack #1 \rbrack}
\def\smno{\smallskip\noindent}
\def\meno{\medskip\noindent}

\def\pano{\par\noindent}

\def\ca#1{{\cal #1}}

\def\Dt{\Delta}
\def\la{\langle}
\def\ra{\rangle}

\def\ts{\textstyle}
\def\cl{\centerline}
\def\pt{\partial}

\def\BZT{{\rm Z{\hbox to 3pt{\hss\rm Z}}}}
\def\BZS{{\hbox{\sevenrm Z{\hbox to 2.3pt{\hss\sevenrm Z}}}}}
\def\BZSS{{\hbox{\fiverm Z{\hbox to 1.8pt{\hss\fiverm Z}}}}}
\def\BQT{\,\hbox{\hbox to -2.8pt{\vrule height 6.5pt width .2pt\hss}\rm Q}}

\def\BQS{\,\hbox{\hbox to -2.1pt{\vrule height 4.5pt width .2pt\hss}$
    \scriptstyle\rm Q$}}
\def\BQSS{\,\hbox{\hbox to -1.8pt{\vrule height 3pt width
    .2pt\hss}$\scriptscriptstyle \rm Q$}}

\def\BCT{\,\hbox{\hbox to -3pt{\vrule height 6.5pt width .2pt\hss}\rm C}}
\def\BCS{\,\hbox{\hbox to -2.2pt{\vrule height 4.5pt width .2pt\hss}$
    \scriptstyle\rm C$}}
\def\BCSS{\,\hbox{\hbox to -2pt{\vrule height 3.3pt width
    .2pt\hss}$\scriptscriptstyle \rm C$}}

\def\section#1{\leftline{\bf\gro #1}\vskip-7pt\line{\hrulefill}}
\def\chapter#1{\leftline{\bf\mit #1}\vskip-7pt\line{\hrulefill}}
\def\unchapter#1{\leftline{\bf #1}\vskip-7pt\line{\hrulefill}}
\def\bibitem#1{\parindent=8mm\item{\hbox to 6 mm{$\q{#1}$\hfill}}}
\font\HUGE=cmbx12 scaled \magstep4
\font\Huge=cmbx10 scaled \magstep4
\font\Large=cmr12 scaled \magstep3

%
%
\nopagenumbers
\rightline{BONN-TH-94-12}
\rightline{hep-th/9408082}
\rightline{August 1994}
\vskip 1cm
\centerline{\HUGE Universit\"at Bonn}
\vskip 10pt
\centerline{\Huge Physikalisches Institut}
\vskip 2cm
\centerline{\Large Extension of the N=2 Virasoro algebra by}
\centerline{\Large two primary fields of dimension 2 and 3}
\vskip 1cm
\centerline{{Ralph Blumenhagen${}^*$}\ \
		   and\ \ {Andreas Wi{\ss}kirchen${}^{**}$}}
\vskip 1cm
\centerline{\bf Abstract}
\mn
We explicitly construct the extension of the $\ts{N=2}$ super Virasoro
algebra by two super primary fields of dimension two and three with vanishing
$u(1)$-charge. Using a super covariant formalism we obtain two different
solutions both consistent for generic values of the central charge $c$. The
first one can be identified with the super ${\cal W}_4$-algebra - the
symmetry algebra of the CP($3$) Kazama-Suzuki model. With the help of
unitarity arguments we predict the self-coupling constant of the field of
dimension two for all super ${\cal W}_n$-algebras. The second solution is
special in the sense that it does not have a finite classical limit
$c\rightarrow\infty$ and generic null fields appear. In the spirit of recent
results in the $\ts{N=0}$ case it can be understood as a unifying $\ts{N=2}$
super ${\cal W}$-algebra for all CP($n$) coset models. It does not admit any
unitary representation.
\vskip 1cm
\pano
Post address:
\pano
Nu{\ss}allee 12
\pano
D-53115 Bonn
\pano
Germany
\pano
e-mail:
\pano
${}^*$ralph@avzw01.physik.uni-bonn.de
\pano
${}^{**}$wisskirc@avzw01.physik.uni-bonn.de
\vfill
\eject
\pageno=1
\footline{\hss\tenrm\folio\hss}
\rm
\chapter{1 Introduction}
\meno
In 1989, motivated by the work of Gepner [11], Kazama and Suzuki [20,21] were
looking for $N=2$ theories having minimal and unitary representations for
$c=9$. They constructed and classified coset models of $N=1$ Kac-Moody
algebras which have an extended $N=2$ supersymmetry. The symmetry algebras of
these models should be generically existing $N=2$ super ${\cal W}$-algebras
[15,23]. Recently, another occurrence of $N=2$ super ${\cal W}$-algebras was
established by Bershadsky et al.\ [4], realizing that the ghost-sector of
${\cal W}_n$-gravity theories coupled to matter contains $N=2$ super
${\cal W}$-algebras.
\pano
Ito [15,16] found via quantum hamiltonian reduction based on the affine super
Lie algebra ${\cal A}(n,n-1)$ that the symmetry algebra of the CP($n-1$)
Kazama-Suzuki model should be the super ${\cal W}_{n}$-algebra, an extension
of the $N=2$ super Virasoro algebra by bosonic superfields of dimension $2$
up to $n-1$. The explicit form of the classical super ${\cal W}_3$-algebra
has been found by Lu et al.\ [22] via Polyakov construction. Later, using
conformal bootstrap techniques, Romans [26] explicitly constructed its
quantum version which has been the only generically existing nonlinear
quantum $N=2$ super ${\cal W}$-algebra constructed up to now. This algebra
was investigated further, classical [19] and quantum [1,17,18] free field
realizations were found. Also higher classical super extensions of the
${\cal W}_3$-algebra [30] were constructed, like the $N=4$ super
${\cal W}_3$-algebra [25] by the 'dual formalism'\ [24]. Inami et al.\ [14]
constructed the extension of the $N=2$ super Virasoro algebra by a superfield
of dimension ${\ts{3\over2}}$. In [6] extensions by superfields of dimension
higher than two were investigated systematically, yielding only
non-deformable solutions.
\pano
Our motivation in this letter was to extend the super ${\cal W}_n$-series.
Besides the expected super ${\cal W}_4$ solution we found a second one which
can be regarded as a 'unifying'\ [7,8] super ${\cal W}$-algebra for the
CP($n$) Kazama-Suzuki models.
\pano
This letter is organized as follows: In the second section we review an $N=2$
supersymmetric manifestly covariant formalism. In the third section we give
the algorithm for construction of the super ${\cal W}_4$-algebra finding two
solutions to be consistent for generic value of the central charge $c$. In
the fourth section we present and discuss the two solutions giving a short
report on unifying algebras. In the fifth section we conclude this letter
presenting a short summary.
\meno
\chapter{2 Holomorphic N=2 supersymmetric CFT}
\meno
We present some formulae about the general structure of holomorphic $N=2$
superconformal field theories [6,27]: using the definitions
$$\ts{Z_{12}= z_1-z_2-
 {\theta_1^+\theta_2^-+\theta_1^-\theta_2^+\over2}\,,
      {\theta}_{12}^\pm={\theta}_1^\pm-{\theta}_2^\pm\,,
 D^\pm = \partial_{{\theta}^\pm} +{1\over 2}
{\theta}^\mp\partial_z}\eqno(2.1)$$
the $N=2$ super Virasoro field is given by:
$$\ts{{\cal L}(Z) = J(z)+{\theta^+G^-(z)
    -{\theta^-}G^+(z)\over\sqrt 2}+\theta^+{\theta^-}L(z)}. \eqno(2.2)$$
An $N=2$ holomorphic superfield of super conformal dimension $\Dt$ and
$u(1)$-charge $Q$ is denoted by:
$$\ts{\Phi_\Dt^Q(Z) = \varphi(z)+{\theta^+{\psi^-}(z)
    -{\theta}^-\psi^+(z)\over\sqrt 2}+\theta^+{\theta^-}\chi(z)
\equiv\Phi_1(z)+{\theta^+{\Phi_2}(z)
    -{\theta}^-\Phi_3(z)\over\sqrt 2}
    +\theta^+{\theta^-}\Phi_4(z)}.\eqno(2.3)$$
The two-point function reads:
$$ \la\Phi_{\Dt_i}^{Q_i} (Z_1)\Phi_{\Dt_j}^{Q_j}(Z_2)\ra=
    {D_{ij}\over Z_{12}^{2\Dt_i}}\left(1-{Q_i\over 2}{\theta_{12}^+
    \theta_{12}^-\over  Z_{12}} \right) \delta_{\Dt_i,\Dt_j}
      \delta_{Q_i,-Q_j}\eqno(2.4)$$
with some constant $D_{ij}$.
\pano
The three-point function is more complicated since several cases have to be
distinguished:
\smno
\item{(i)} If $Q_i+Q_j=Q_k$  then
$$\ts{\la \Phi_{\Dt_k}^{-Q_k}(Z_1)\>\Phi_{\Dt_i}^{Q_i}(Z_2)\>
\Phi_{\Dt_j}^{Q_j}(Z_3)\ra  =
C_{ijk}{2+\gamma_1\alpha_{ijk}\Theta_{ijk}^+\Theta_{ijk}^-
\over Z_{12}^{\gamma_1}\> Z_{23}^{\gamma_2} \> Z_{13}^{\gamma_3}}
\prod_{m<n}(1+{Q_n-Q_m\over6}{\theta_{mn}^+\theta_{mn}^-\over Z_{mn}})}.
\eqno (2.5) $$
\item{(ii)} If $Q_i+Q_j=Q_k\pm1$  then
$$\ts{\la \Phi_{\Dt_k}^{-Q_k}(Z_1)\>\Phi_{\Dt_i}^{Q_i}(Z_2)\>
\Phi_{\Dt_j}^{Q_j}(Z_3)\ra  =
{C_{ijk}\Theta_{ijk}^\pm
\over Z_{12}^{\gamma_1}\> Z_{23}^{\gamma_2} \> Z_{13}^{\gamma_3}}
(1+\sum_{m<n}{Q_n-Q_m\over6}{\theta_{mn}^+\theta_{mn}^-\over Z_{mn}})}
\eqno (2.6) $$
where $\gamma_1=\Dt_k+\Dt_i-\Dt_j$,  $\gamma_2=\Dt_i+\Dt_j-\Dt_k$
and $\gamma_3=\Dt_k+\Dt_j-\Dt_i$.
$C_{ijk}$ and $\alpha_{ijk}$ are {\it independent} parameters.
$$\ts{\Theta_{ijk}^\pm={Z_{ij}\theta_k^\pm+Z_{jk}\theta_i^\pm+Z_{ki}
		    \theta_j^\pm
		    +{1\over6}
		   \left(\theta_i^\pm\theta_{jk}^\mp
		    (\theta_j^\pm+\theta_k^\pm)
		   +\theta_j^\pm\theta_{ki}^\mp
		    (\theta_k^\pm+\theta_i^\pm)
		   +\theta_k^\pm\theta_{ij}^\mp
		    (\theta_i^\pm+\theta_j^\pm)\right)
		    \over\sqrt{Z_{ij}Z_{jk}Z_{ki}}}} \eqno(2.7)$$
denotes the pair of Grassmann odd $N=2$ super M\"obius ($\equiv Osp(2,2)$)
invariants [27].
\smno
Following [6], the coupling constants $\ts{C_{ij}^k}$ occurring in the OPE
are determined by the linear system $\ts{C_{ij}^lD_{lk}=C_{ijk}}$.
\pano
For our calculations we use a different but equivalent form of the
three-point function which is not invariant under permutations.
\pano
Naive normal ordered products (NOPs), defined naturally by terms of the
regular part of the OPE, are not quasiprimary, but one can project them
onto their quasiprimary component ${\cal N}_s$, e.g.
$$\ts{{\cal N}_s({\cal L}{\cal L})=N_s({\cal L}{\cal L})
	   -{1\over 3}[D^-,D^+]{\cal L}}. \eqno(2.8)$$
\meno
{\chapter{3 Algorithm}}
\meno
We will apply the manifestly covariant formalism presented in [6] to the
construction of an $\ts{N=2}$ super ${\cal W}$-algebra with {\it three}
generators, ${\cal SW}(1,2,3)$: We extend the $\ts{N=2}$ super Virasoro
algebra by two additional super primary fields
$\ts{{\cal W}(\Delta\hskip -1pt =\hskip -1pt 2,
Q\hskip -1pt =\hskip -1pt 0)}$ and
$\ts{{\cal V}(\Delta\hskip -1pt =\hskip -1pt 3,
Q\hskip -1pt =\hskip -1pt 0)}$.
Note that it would be almost impossible to directly
construct an $\ts{N=0}$ ${\cal W}$-algebra containing twelve
generators. Only the $\ts{N=2}$ structure makes such a calculation
practicable.
\pano
We choose the following normalization of the two-point functions:
$$\ts{{\la{\cal W}(Z_1){\cal W}(Z_2)\ra={{c/2}\over{Z_{12}^4}}\,,\quad\quad
  \la{\cal V}(Z_1){\cal V}(Z_2)\ra={{c/3}\over{Z_{12}^6}}.
  }}\eqno(3.1) $$
For explicit calculations we expand the components of the superfields into
Fourier modes, making
it possible to apply Lie algebra methods such as the universal polynomials
[5]. We use $Mathematica^{TM}$[29] and the C-program {\it 'commute'}\ [12].
For consistency of the algebra all commutators have to satisfy Jacobi
identities leading to restrictions on the self-couplings and
the central charge $c$. Thus we proceed as follows:
\meno
$\bullet$
We write down the most general ans\"atze for the super OPEs
schematically given by:
$$\eqalignno{ {\cal W}\circ{\cal W}&=[\ca L]+[\ca W]+[\ca V]
		  &\cr
	      {\cal V}\circ{\cal W}&=[\ca L]+[\ca W]+[\ca V]
		  +[\cal WW]            &(3.2)\cr
	      {\cal V}\circ{\cal V}&=[\ca L]+[\ca W]+[\ca V]
		  +[\cal WW]+ [\cal VW].   &\cr}$$
For each dimension occurring in the OPE we need a basis of super quasiprimary
normal ordered products. All relevant NOPs are shown in table 1.
\meno
\cl{\vbox{
\hbox{\vbox{\offinterlineskip
\def\tablespace{height2pt&\omit&&\omit&&\omit&\cr}
\def\tablerule{\tablespace\noalign{\hrule}\tablespace}
\hrule\halign{&\vrule#&\strut\hskip0.2cm\hfil#\hfill\hskip0.2cm\cr
\tablespace
& $\Dt$ && $Q$ && ${\rm NOPs}\ \ca F$ &\cr
\tablerule\tablerule
& $1$ && $0$ && ${\cal L}$ & \cr
\tablespace
& $2$ && $0$ && ${\cal N}_s({\cal LL}) \;; {\cal W}$ & \cr
\tablespace
& $3$ && $0$ && ${\cal N}_s({\cal N}_s({\cal LL}){\cal L}) \,,
     {\cal N}_s({\cal L}[D^-,D^+]{\cal L}) \;;
     {\cal N}_s({\cal WL}) \;;
     {\cal V}$ & \cr
\tablespace
& $7\over2$ && $\mp1$ && ${\cal N}_s({\cal L}{D}^\pm\pt{\cal L}) \;;
   {\cal N}_s({\cal W}{D}^\pm{\cal L})$ & \cr
\tablespace
& $4$ && $0$ && ${\cal N}_s({\cal N}_s({\cal N}_s({\cal LL})
	       {\cal L}){\cal L}) \,,
 {\cal N}_s({\cal N}_s({\cal L}[D^-,D^+]{\cal L}){\cal L}) \,,
 {\cal N}_s({\cal L}\pt^2{\cal L}) \;; $ & \cr
\tablespace
&  &&  && $
 {\cal N}_s({\cal N}_s({\cal WL}){\cal L}) \,,
 {\cal N}_s({\cal W}[D^-,D^+]{\cal L}) \,,
 {\cal N}_s({\cal W}\pt{\cal L}) \;;
 {\cal N}_s({\cal VL}) \;;
 {\cal N}_s({\cal WW})$ & \cr
\tablespace
& $9\over2$ && $\mp1$ && ${\cal N}_s({\cal N}_s({\cal L}D^\pm\pt
	       {\cal L}){\cal L}) \;;
 {\cal N}_s({\cal N}_s({\cal W}D^\pm{\cal L}){\cal L}) \,,
 {\cal N}_s({\cal W}D^\pm\pt{\cal L}) \;;
 {\cal N}_s({\cal V}D^\pm{\cal L})$ &\cr
\tablespace
& $5$ && $0$ &&
${\cal N}_s({\cal N}_s({\cal N}_s({\cal N}_s({\cal LL})
	       {\cal L}){\cal L}){\cal L}) \,,
 {\cal N}_s({\cal N}_s({\cal N}_s({\cal L}
  [D^-,D^+]{\cal L}){\cal L}){\cal L}) \,, $
 & \cr
\tablespace & && && $
 {\cal N}_s({\cal N}_s({\cal L}\pt^2{\cal L}){\cal L}) \,,
 {\cal N}_s({\cal L}[D^-,D^+]\pt^2{\cal L}) \,,
 {\cal N}_s({\cal N}_s({\cal L}D^\pm\pt{\cal L})D^\mp{\cal L}) \;;
 $ & \cr
\tablespace & && && $
 {\cal N}_s({\cal N}_s({\cal N}_s({\cal WL})
	       {\cal L}){\cal L}) \,,
 {\cal N}_s({\cal N}_s({\cal W}[D^-,D^+]{\cal L}){\cal L}) \,,
 {\cal N}_s({\cal W}[D^-,D^+]\pt{\cal L}) \,,
$ & \cr
\tablespace
&  &&  && $
 {\cal N}_s({\cal N}_s({\cal W}D^\pm{\cal L})D^\mp{\cal L}) \,,
 {\cal N}_s({\cal W}\pt^2{\cal L}) \;;
 {\cal N}_s({\cal N}_s({\cal VL}){\cal L}) \,,
 {\cal N}_s({\cal V}[D^-,D^+]{\cal L}) \,,
 $ & \cr
\tablespace
&  &&  && $
 {\cal N}_s({\cal V}\pt{\cal L}) \;;
 {\cal N}_s({\cal N}_s({\cal WW}){\cal L}) \,,
 {\cal N}_s({\cal W}[D^-,D^+]{\cal W}) \;;
 {\cal N}_s({\cal VW})  $ & \cr
\tablespace
& $11\over2$ && $\mp1$ && $
 {\cal N}_s({\cal N}_s({\cal N}_s({\cal N}_s({\cal LL})
	       {\cal L}){\cal L})D^\pm{\cal L}) \,,
 {\cal N}_s({\cal N}_s({\cal N}_s({\cal L}[D^-,D^+]{\cal L})
   {\cal L})D^\pm{\cal L}) \,, $ & \cr
\tablespace
& && &&
 ${\cal N}_s({\cal N}_s({\cal N}_s({\cal L}
	     D^\pm\pt{\cal L}){\cal L}){\cal L}) \,,
 {\cal N}_s({\cal N}_s({\cal L}\pt^2{\cal L})D^\pm{\cal L}) \;;
 {\cal N}_s({\cal N}_s({\cal W}\pt{\cal L})D^\pm{\cal L}) \,,  $ &\cr
\tablespace
&  &&  && ${\cal N}_s({\cal N}_s({\cal N}_s({\cal W}D^\pm
	       {\cal L}){\cal L}){\cal L}) \,,
 {\cal N}_s({\cal N}_s({\cal W}[D^-,D^+]{\cal L})D^\pm{\cal L}) \,,
 {\cal N}_s({\cal W}D^\pm\pt^2{\cal L}) \,, $ &\cr
\tablespace
& && && ${\cal N}_s({\cal N}_s({\cal W}D^\pm\pt{\cal L}){\cal L}) \;;
      {\cal N}_s({\cal N}_s({\cal V}D^\pm{\cal L})
	     {\cal L}) \,,
     {\cal N}_s({\cal V}D^\pm\pt{\cal L}) \;;
 {\cal N}_s({\cal W}D^\pm\pt{\cal W}) \,, $ & \cr
\tablespace
& && && ${\cal N}_s({\cal N}_s({\cal WW})D^\pm{\cal L}) \;;
	   {\cal N}_s({\cal V}D^\pm{\cal W})$ & \cr
\tablespace}\hrule}}
\hbox{\hskip 0.5cm Table 1:\hskip 0.5cm Quasiprimary NOPs up to dimension
 $\ts{11\over2}$ }}}
\smno
The Kac determinants in the vacuum sector are presented in table 2.
\smno
\cl{\vbox{
\hbox{\vbox{\offinterlineskip
\def\tablespace{height2pt&\omit&&\omit&\cr}
\def\tablerule{\tablespace\noalign{\hrule}\tablespace}
\hrule\halign{&\vrule#&\strut\hskip0.2cm\hfil#\hfill\hskip0.2cm\cr
\tablespace
& $\Dt$ && ${\rm det}(D_{\Dt})\sim$  &\cr
\tablerule\tablerule
& $1$ &&  $c$ &\cr
\tablespace
& $2$ &&  $c^2(c-1)$ &\cr
\tablespace
& $3$ &&  $c^4(c-1)(c+6)(2c-3)(5c-12)$ &\cr
\tablespace
& ${\ts{7\over2}}$ &&  $c^4{(c-1)}^2{(c+3)}^2$ &\cr
\tablespace
& ${\ts{9\over2}}$ &&  $c^8{(c-1)}^4{(c+3)}^2{(c+6)}^2{(2c-3)}^2{(5c-12)}^2$
&\cr
\tablespace}\hrule}}
\hbox{\hskip 0.5cm Table 2: \hskip 0.5cm Kac determinants}}}
\smno
We omit the levels $4$, $5$ and ${\textstyle{11\over2}}$ since these
determinants contain a priori unknown self-couplings. They will be
tabulated below after having presented the solutions obtained from Jacobi
identities.
\pano
Although three-point functions involving ${\cal V}$, ${\cal W}$ and
${\cal F}$ vanish identically for all fields ${\cal F}\in[{\cal L}]$, it is
necessary to assume the appearance of $[{\cal L}]$ in the OPE ${\cal V}\circ
{\cal W}$ because the $D$-matrices $(D_{ij})$ are not blockdiagonal with
respect to the
conformal families, whenever $[{\cal WW}]$ or $[{\cal VW}]$ appear.
\meno
$\bullet$
The next step is to expand the fields into their components and to evaluate
the structure constants for the lowest components. We obtain the super
coupling
constants by which all structure constants for the higher components are
determined [6]. Note that $\ts{N=2}$
supersymmetry implies that we only have to
calculate one $D$-matrix for every super-conformal dimension and one or
two coupling constants
per field (there are about 30 structure constants of the components!). In
this step covariance reduces the effort considerably.
\meno
$\bullet$
Finally, we have to check Jacobi identities. It is sufficient to check them
only for the additional primaries because Jacobi identities involving the
super Virasoro field are satisfied automatically by $Osp(2,2)$-invariance.
Let $\ts{(\phi_i\phi_j\phi_k\phi_l)}$ be the Jacobi identity equivalent to
the associativity of the four point function involving these four
fields.
Exploiting symmetries such as charge conjugation, there are only eight types
of Jacobi identities for the component fields of one superfield with
$u(1)$-charge $Q=0$:
$$ (\varphi\hskip 0.04cm\varphi\hskip 0.04cm\varphi\hskip 0.04cm\varphi) \,,
   (\varphi\hskip 0.04cm\varphi\hskip 0.04cm\varphi\hskip 0.04cm\chi)   \,,
   (\varphi\hskip 0.04cm\varphi\hskip 0.04cm\chi\hskip 0.04cm\chi) \,,
   (\varphi\hskip 0.04cm\chi\hskip 0.04cm\chi\hskip 0.04cm\chi) \,,
   (\chi\hskip 0.04cm\chi\hskip 0.04cm\chi\hskip 0.04cm\chi) \,,
   (\varphi\hskip 0.04cm\varphi\hskip 0.04cm\psi^+\psi^-) \,,
   (\psi^+\psi^-\chi\hskip 0.04cm\chi) \,,
   (\psi^+\psi^-\psi^+\psi^-).\eqno(3.3)$$
Altogether we have to check $48$ Jacobi identities, but only four of them
turn out to be independent.
\meno
\chapter{4 Results}
\meno
In this section we present the results of our calculations. We note
beforehand that we found two and only two solutions existing for generic
value of $c$ which we denote by ${{\cal SW}(1,2,3)}^{[{\rm I,II}]}$.
The OPE ${\cal W}\circ{\cal W}$ is given in table 3.
\meno
\cl{\vbox{
\hbox{\vbox{\offinterlineskip
\def\tablespace{height2pt&\omit&&\omit&&\omit&\cr}
\def\tablerule{\tablespace\noalign{\hrule}\tablespace}
\hrule\halign{&\vrule#&\strut\hskip0.2cm\hfil#\hfill\hskip0.2cm\cr
\tablespace
& ${\rm NOP}\ \ca F$ && $C_{\cal WW}^{\ca F}$ && $C_{\cal WW}^{\ca F}
\alpha_{\cal WWF}$ &\cr
\tablerule\tablerule
& $\ca L$ && $0$ && $3$ & \cr
\tablespace
& ${\cal N}_s({\cal L}{\cal L})$ && $-{3\over 2(c-1)}$ && $0$ &\cr
\tablespace
& ${\cal N}_s({\cal N}_s({\cal L}{\cal L}){\cal L})$ && $0$ &&
 $-{3(4c+3)\over (2c-3)(c-1)(c+6)}$ &\cr
\tablespace
& ${\cal N}_s({\cal L}[D^-,D^+]{\cal L})$ && $0$ &&
  ${23c-30\over 2(2c-3)(c+6)}$&\cr
\tablespace
& ${\cal N}_s({\cal L}D^\pm\pt{\ca L})$ && $-{6\over c-1}$ && $-$ &\cr
\tablespace
& ${\cal V}$ && $0$ && ${3\over2}C_{\cal VW}^{\cal W}\alpha_{\cal VWW}$ &\cr
\tablespace
& ${\cal W}$ && $C_{\cal WW}^{\cal W}$ && $0$ &\cr
\tablespace
& ${\cal N}_s({\cal WL})$ &&  $0$ && ${14 C_{\cal WW}^{\cal W}\over 5c-12}$
&\cr
\tablespace
& ${\cal N}_s({\cal W}D^\pm{\ca L})$ &&
$\pm{12C_{\cal WW}^{\cal W}\over c+3}$
&& $-$ &\cr
\tablespace}\hrule}}
\hbox{\hskip 0.5cm Table 3:\hskip 0.5cm Structure constants of the OPE
${\cal W}\circ{\cal W}$
}}}
\meno
We skip the presentation of the other structure constants since they are
complicated rational functions in $c$ and the couplings among the primaries
themselves. They simplify only considerably after inserting the results
obtained from checking Jacobi identities. This also fixes the
values of the self-couplings appearing in table 3.
A complete list of structure constants can be found in [28].
\meno
\unchapter{4.1 ${{\cal SW}(1,2,3)}^{[{\rm I}]}$,
the super ${\cal W}_4$-algebra}
\meno
We obtain a first solution for which the four free couplings in the ansatz
(3.2) are:
$$ \eqalignno{ \left(C_{\cal WW}^{\cal W}\right)^2&={(c+3)^2(7c-27)^2
   \over    2(1-c)(c-21)(c+9)(5c-9) } &\cr
    \left(C_{\cal VW}^{\cal W}\alpha_{\cal VWW}\right)^2&
     ={36(c-33)(c-3)(c-1)^2(c+3)(c+18)
    \over  (c-21)(c+6)(c+9)(2c-3)(5c-12)(5c-9) } &(4.1.1)\cr
    \left(C_{\cal VW}^{\cal V}\right)^2&={9(c+6)^2(2c-3)^2(7c-27)^2
    \over  2(9-5c)(c-21)(c-1)(c+9)(5c-12)^2 } &\cr
    \left(C_{\cal VV}^{\cal V}\alpha_{\cal VVV}\right)^2&=
   &\cr
   &  { 4\left( 1312200-1234926c
    +393093c^2-72306c^3+7641c^4+123c^5+7c^6\right)^2 \over
   (c-33)(c-21)(c-3)(c+3)(c+6)(c+9)(c+18)(2c-3)(5c-12)^3(5c-9)}.
    &\cr}$$
The remaining Kac determinants now read as in table 4.
\smno
\cl{\vbox{
\hbox{\vbox{\offinterlineskip
\def\tablespace{height2pt&\omit&&\omit&\cr}
\def\tablerule{\tablespace\noalign{\hrule}\tablespace}
\hrule\halign{&\vrule#&\strut\hskip0.2cm\hfil#\hfill\hskip0.2cm\cr
\tablespace
& $\Dt$ && ${\rm det}(D_{\Dt})\sim$  &\cr
\tablerule\tablerule
& $4$ && $c^8(c-33)(c-3){(c-1)}^5(c+3)(c+6)(c+18)(2c-3)
	(5c-12)(5c+27)(7c-27)/$  &\cr
\tablespace
& && $\left((c-21)(c+9)\right)$  &\cr
\tablespace
& $5$ && $c^{18}(c-57)(c-33)(c-18){(c-3)}^2{(c-1)}^{11}{(c+3)}^4
   {(c+6)}^4(c+15){(c+18)}^2(c+36)$
   & \cr
\tablespace
&  && $(2c-9)
  {(2c-3)}^4{(5c-12)}^4{(5c+27)}^2{(7c-27)}^2/\left({(c-21)}^3
  {(c+9)}^3{(5c-9)}^2\right)$
 & \cr
\tablespace
& ${\ts{11\over2}}$ &&
$c^{28}{(c-33)}^2{(c-27)}^2{(c-18)}^2
  {(c-3)}^2{(c-1)}^{20}{(c+3)}^{10}{(c+6)}^6{(c+15)}^2
 {(c+18)}^2$
 &\cr
\tablespace
&  &&
 ${(2c-3)}^6{(5c-12)}^6{(5c+27)}^4{(7c-27)}^2/
 \left({(c-21)}^6{(c+9)}^6{(5c-9)}^4\right) $
 &\cr
\tablespace}\hrule}}
\hbox{\hskip 0.5cm Table 4: \hskip 0.5cm Kac determinants [I]
}}}
\smno
In contrast to the second solution this algebra has a finite classical
limit $c\rightarrow\infty$ which is a hint
that only this solution can be the super ${\cal W}_4$-algebra. In
order to support this assumption, we use the decomposition of CP($n$)
Kazama-Suzuki models into three bosonic coset models (see [26] and
references therein):
$$  {\wh{su}(n+1)_k\oplus \wh{so}(2n)_1\over  \wh{su}(n)_{k+1}\oplus
    \wh{u}(1)}\cong
   {\wh{su}(k)_n\oplus \wh{su}(k)_1\over  \wh{su}(k)_{n+1}}
    \oplus
   {\wh{su}(n)_k \oplus \wh{su}(n)_1\over \wh{su}(n)_{k+1} }
   \oplus \wh{u}(1). \eqno(4.1.2)$$
The equivalence has to be understood as up to questions of
finite reducibility of the representation theory [10].
\pano
Using the T-equivalence [2,9]
$$ {\wh{su}(k)_n\oplus \wh{su}(k)_1\over  \wh{su}(k)_{n+1}}\cong
   {\wh{su}(n+1)_k \over \wh{su}(n)_{k}\oplus \wh{u}(1)}\eqno(4.1.3)$$
we can substitute the first coset on the r.h.s.\ of (4.1.2) by a bosonic
CP($n$) coset. For $\ts{n=1}$ one can rigorously prove a dual
equivalence [2].
\pano
Suppose ${{\cal SW}(1,2,3)}^{\rm [I]}$ to be the symmetry algebra of the
CP($3$) Kazama-Suzuki model. Then the decomposition (4.1.2) will induce a
transformation of the subalgebra
of the fields of dimension two, namely $N(JJ),L,\varphi={\cal W}_1$
into a direct sum of three Virasoro algebras. These three Virasoro fields
can be constructed explicitly:
Denoting
 $$   \gamma=\sqrt{4c-\left( C_{\cal WW}^{\cal W}\right)^2+
    c \left(C_{\cal WW}^{\cal W}\right)^2 } \eqno(4.1.4)$$
we obtain three mutually commuting Virasoro generators with their
corresponding central charges:
\pano
$\bullet$ bosonic CP$(n)$-part
$$  T_1={\ts {\gamma+\sqrt{c-1}C_{\cal WW}^{\cal W} \over 2\gamma}
     }L-{\ts {\sqrt{c-1}\over \gamma}}\varphi-{\ts {3\left(\gamma+\sqrt{c-1}
     C_{\cal WW}^{\cal W}\right)\over 4c\gamma}}
       N(JJ) \eqno(4.1.4{\rm a})$$
$$  \tilde c_1=(c-1){\gamma+\sqrt{c-1}C_{\cal WW}^{\cal W}
	\over 2\gamma} ,$$
$\bullet$ $\ca W_n$-part
$$  T_2={\ts {\gamma-\sqrt{c-1}C_{\cal WW}^{\cal W} \over 2\gamma}
     }L+{\ts {\sqrt{c-1}\over \gamma}}\varphi-{\ts {3\left(\gamma-\sqrt{c-1}
     C_{\cal WW}^{\cal W}\right)\over 4c\gamma}}
       N(JJ) \eqno(4.1.4{\rm b})$$
$$  \tilde c_2=(c-1){\gamma-\sqrt{c-1}C_{\cal WW}^{\cal W}
	   \over 2\gamma} ,$$
$\bullet$ $\wh{u}(1)$-part
$$  T_3={\ts {3\over 2c}}N(JJ) \eqno(4.1.4{\rm c})$$
$$  \tilde c_3=1.$$
Of course, the self-coupling $C_{\cal WW}^{\cal W}$ depends on the super
${\cal W}$-algebra. Inserting the solution $(4.1.1)$ yields the central
charges:
$$\eqalignno{ \tilde c_1&= {(c+9)(5c-9)\over 3(c+27)} &\cr
	      \tilde c_2&= {2c(21-c)\over 3(c+27)} &(4.1.5)\cr
	      \tilde c_3&= 1. &\cr }$$
Unitary representations of ${{\cal SW}(1,2,3)}^{\rm [I]}$ are also unitary
representations of the ${\cal W}_3$-part.
Therefore, $$\tilde c_2=2\left(1-{12\over (k+3)(k+4)}\right) \eqno(4.1.6) $$
is a necessary condition for $c$ of the unitary series of
${{\cal SW}(1,2,3)}^{\rm [I]}$. Plugging (4.1.6) in (4.1.5) we get a
quadratic equation. Its solution consists of one ascending and one descending
branch of central charges:
$$\eqalignno{  c&={9k\over k+4}, \quad\quad
	       c={9(k+7)\over (k+7)-4}. &(4.1.7)\cr }$$
The first branch is actually the unitary series of the CP($3$) Kazama-Suzuki
model, confirming the identification of ${{\cal SW}(1,2,3)}^{\rm [I]}$ as its
symmetry algebra - the super ${{\cal W}_4}$-algebra.
\pano
The second series can be formally obtained from the first one by the
substitution $k\rightarrow -k-7$. It
is related to non-compact coset models (see ref.\ [3] for more details).
\pano
Furthermore, we have looked for primary fields in the different summands of
(4.1.2).
Actually, in
the ${\cal W}_3$-part there is a spin-$3$ primary with respect to $T_2$
generating a ${\cal W}(2,3)$ together with $T_2$ - we have checked that there
is no spin-$4$ generator in this part.
Concerning the investigation of the bosonic CP($n$) models in ref.\ [8]
there should be a ${\cal W}(2,3,\dots,19)$ in the CP($3$)-part - we
constructed the fields of dimension $3$ and $4$
explicitly. Those three fields are presented in the appendix (A.1-A.3).
\pano
The formulae (4.1.3-4.1.4) do not depend on the special structure of
${\cal SW}(1,2,3)$.
Reversing the whole procedure one can obtain the general self-coupling
constant of the field of dimension two
for any super $\ca{W}_{n+1}$-algebra. To this end, we require $c$ and
$\tilde c_2$ to take values in the unitary series of the CP($n$)
Kazama-Suzuki model and the ${\cal W}_n$ model, respectively:
$$ c={3nk\over k+n+1}, \quad\quad
   \tilde c_2=(n-1)\left(1-{n(n+1)\over (k+n)(k+n+1)}\right) \eqno(4.1.8) $$
Inserting $\tilde c_2$ in (4.1.4b) and eliminating $k$ by $c$ yields:
$$   \left( C_{\cal WW}^{\cal W}\right)^2=
   {(c+3)^2\left( (2n+1)c-3{n}^2\right)^2\over
     (n-1)\,(c-1)\,\left(c+3n\right)\,(6n+3-c)\,\left((n+2)c-3n\right)
     }.\eqno(4.1.9)$$
In the following we will use this formula for the identification of the
second solution
$\ca{SW}(1,2,3)^{[{\rm II}]}$.
\meno
\unchapter{4.2 ${{\cal SW}(1,2,3)}^{[{\rm II}]}$,
    a unifying super ${\cal W}$-algebra}
\meno
In order to identify the second solution,
we summarize the most important results
about $N=0$ unifying ${\cal W}$-algebras [7,8].
\pano
Unifying ${\cal W}$-algebras interpolate the rank $n$ of Casimir algebras
${\cal WL}_n$ at particular values of the central charge.
Using the coset realizations of these algebras, the unifying
${\cal W}$-algebras can be regarded as a generalization
of level-rank-duality [2,9]. These identifications
between a priori different ${\cal W}$-algebras for particular values of the
central charge are closely related to the fact that for certain values
of the central charge some generators become null fields leading to a
'truncation' of the ${\cal W}$-algebra.
Another important feature of unifying algebras is that certain fields which
one would expect to appear in the singular part of the OPE of the primaries
are null fields. Finally, the absence of a finite classical limit
$c\rightarrow\infty$ is characteristic for
unifying ${\cal W}$-algebras in all explicitly known cases.
\meno
The structure constants of the second solution among the primaries are:
$$ \eqalignno{ \left(C_{\cal WW}^{\cal W}\right)^2&={c(c+3)^2 \over
	 2(1-c)(1+c) } &\cr
    \left( C_{\cal VW}^{\cal W}\alpha_{\cal VWW}\right)^2&
       ={8(24-c)(c-1)^2(c+3)^2
    \over  3(c+1)(c+6)(2c-3)(5c-12) } &(4.2.1)\cr
    \left(C_{\cal VW}^{\cal V}\right)^2&={9c(c+6)^2(2c-3)^2
    \over  2(1-c)(1+c)(5c-12)^2 } &\cr
    \left( C_{\cal VV}^{\cal V}\alpha_{\cal VVV}\right)^2&
	={ (24-c)\left( 1296-2484c
    +36c^2+315c^3+74c^4\right)^2 \over
   6(c+1)(c+3)^2(c+6)(2c-3)(5c-12)^3 }.     &\cr}$$
Again the rest of the Kac determinants becomes quite simple (see table 5).
\smno
\cl{\vbox{
\hbox{\vbox{\offinterlineskip
\def\tablespace{height2pt&\omit&&\omit&\cr}
\def\tablerule{\tablespace\noalign{\hrule}\tablespace}
\hrule\halign{&\vrule#&\strut\hskip0.2cm\hfil#\hfill\hskip0.2cm\cr
\tablespace
& $\Dt$ && ${\rm det}(D_{\Dt})\sim$  &\cr
\tablerule\tablerule
& $4$ && $c^9{(c-1)}^5{(c+3)}^2(c+6)(2c-3)(5c-12)(5c+27)$ &\cr
\tablespace
& $5$ && $c^{20}{(c-1)}^{11}{(c+3)}^6{(c+6)}^4{(c+8)}^2{(2c-3)}^4
	       {(5c-12)}^4{(5c+27)}^2/{(c+1)}^2$
 & \cr
\tablespace
& ${\ts{11\over2}}$ && $
 c^{28}{(c-1)}^{18}{(c+3)}^{14}{(c+6)}^4{(c+8)}^2{(2c-3)}^4
 {(5c-12)}^4{(5c+27)}^2/{(c+1)}^2
 $ &\cr
\tablespace}\hrule}}
\hbox{\hskip 0.5cm Table 5: \hskip 0.5cm Kac determinants [II]}}}
\smno
There is a pair of generic null fields of dimension ${\textstyle{11\over2}}$
whose explicit form can be found in the appendix (A.4).
For that reason the determinant at this level is based on only 26 fields,
not on 28 as it is for $\ts{{{\cal SW}(1,2,3)}^{[{\rm I}]}}$.
\pano
The occurrence of those null fields and the non-existing classical limit
$c\rightarrow\infty$ of
the structure constants give rise to the conjecture that this algebra is
a unifying super ${\cal W}$-algebra.
\pano
Inserting the solution (4.2.1) in (4.1.4a-c), the central charge simply
splits in the following way:
$$   c=-2+(c+1)+1, \eqno(4.2.2)$$
corresponding to the decomposition:
$$  {\wh{su}(n+1)_{-{n\over 2}}\oplus \wh{so}(2n)_1\over
   \wh{su}(n)_{-{n-2\over 2}}\oplus
    \wh{u}(1)}\cong
   {\wh{su}(n+1)_{-{n\over 2}} \over \wh{su}(n)_{-{n\over 2}}\oplus
    \wh{u}(1)}\oplus
   {\wh{su}(n)_{-{n\over 2}} \oplus \wh{su}(n)_1\over
    \wh{su}(n)_{-{n-2\over 2} } }
   \oplus \wh{u}(1) \eqno(4.2.3)$$
which we obtain from (4.1.2) by inserting (4.1.3) and formally substituting
$k$ by
$\ts{-{n\over2}}$. Note that the bosonic CP($n$) model at level
$k=-{n\over2}$ always carries the central charge $\tilde c=-2$ [8].
The same substitution applied to the unitary series of the CP($3$)
Kazama-Suzuki model should lead to a truncation of
$\ca{SW}(1,2,3)^{[{\rm II}]}$ to the super $\ca W_{n+1}$-algebra
for $n<3$ and vice versa for $n>3$ (for $n=3$ one has
$\ca{SW}(1,2,3)^{[{\rm II}]}=\ca{SW}(1,2,3)^{[{\rm I}]}$)
at $$  c=-{3n^2\over n+2}. \eqno(4.2.4)$$
This can be shown explicitly for all structure constants of the first three
examples:
\item{$\bullet$} $c=-1$:
All self-coupling constants diverge. After rescaling the fields with a
factor $(c+1)^x$ such
that the structure constants will be finite, ${\cal W}$ and ${\cal V}$ become
null fields.
\item{$\bullet$} $c=-3$:
$C_{\cal VW}^{\cal W}\alpha_{\cal VWW}$ and $C_{\cal VV}^{\cal V}
\alpha_{\cal VVV}$ diverge. The redefined
field ${\cal V}$ will be a null field and
$C_{\cal WW}^{\cal W}$ is equal to the self-coupling constant of
the super ${\cal W}_3$-algebra [26].
     \item{$\bullet$} $c=-{27\over5}$:
The structure constants of ${{\cal SW}(1,2,3)}^{[{\rm II}]}$ are equal to the
corresponding ones of ${{\cal SW}(1,2,3)}^{[{\rm I}]}$.
\smno
That reads briefly as in table 6.
\smno
\cl{\vbox{
\hbox{\vbox{\offinterlineskip
\def\tablespace{height2pt&\omit&&\omit&\cr}
\def\tablerule{\tablespace\noalign{\hrule}\tablespace}
\hrule\halign{&\vrule#&\strut\hskip0.2cm\hfil#\hfill\hskip0.2cm\cr
\tablespace
& $c$ && {\rm truncation}  &\cr
\tablerule\tablerule
& $-1$ && $\ca{SW}(1,2,3)^{[{\rm II}]}\to SVir_{N=2}$  & \cr
& $-3$ && $\ca{SW}(1,2,3)^{[{\rm II}]}\to \ca{SW}(1,2)$  & \cr
& $-{27\over 5}$ && $\ca{SW}(1,2,3)^{[{\rm II}]} =
    \ca{SW}(1,2,3)^{[{\rm I}]}$  & \cr
\tablespace}\hrule}}
\hbox{\hskip 0.5cm Table 6: \hskip 0.5cm Truncations}}}
\smno
Furthermore, for $n>3$ $C_{\cal WW}^{\cal W}$ of
$\ca{SW}(1,2,3)^{[{\rm II}]}$
is equal to the structure constant (4.1.9) of the super $\ca W_{n+1}$-algebra
for the corresponding value of $c$, confirming the identification of
$\ca{SW}(1,2,3)^{[{\rm II}]}$ as a unifying super ${\cal W}$-algebra.
\pano
Finally, we have checked that the symmetry algebras of the first and
second part in the decomposition (4.2.3) are a ${\cal W}(2,3)$ with
internal $\tilde c=-2$ and the Virasoro algebra, respectively.
The spin-$3$ field is shown in the appendix (A.5).
\pano
In virtue of $\tilde c=-2$ in the ${\cal W}(2,3)$ symmetry algebra
there is no unitary representation of ${{\cal SW}(1,2,3)}^{[{\rm II}]}$
for any value of $c$: Suppose
${{\cal SW}(1,2,3)}^{[{\rm II}]}$ to have a unitary representation.
Due to the meaning of
equivalence in (4.1.2)
it will be decomposable into
unitary representations of the symmetry algebras of the cosets on the r.h.s.\
of (4.2.3).
However, the ${\cal W}(2,3)$ symmetry algebra will not admit
unitary representations because of its negative central charge.
\meno
\chapter{5 Conclusion and outlook}
\meno
Using a manifestly covariant formalism, we were able to determine the
complete
structure of ${\cal SW}(1,2,3)$. The most important result
is the existence
of {\it two} generic solutions.
\pano
The first one can be identified with the super ${\cal W}_4$-algebra
possessing a finite classical limit $c\rightarrow\infty$.
Using a decomposition of the underlying coset
model - on the algebraic level one obtains three subalgebras - and inserting
their well-known unitary series we recovered the ascending unitary branch of
the CP($3$)
Kazama-Suzuki model. There is a descending second series due to its
non-compact version. Furthermore, we identified the subalgebras and
constructed some of their generators. Inverting
this way of argumentation leads to a prediction of the self-coupling
constant of the field of dimension two for every super ${\cal W}_n$-algebra.
\pano
The second solution has no
finite classical limit $c\rightarrow\infty$.
The decomposition mentioned above
leads to subalgebras with few generators: besides a $\wh{u}(1)$-part
a Virasoro algebra and a ${\cal W}(2,3)$ with
internal $\tilde c=-2$ independent of the overall value of $c$.
Furthermore, the occurrence of a pair of generic null fields strongly
indicates that this solution is a unifying $N=2$ super
${\cal W}$-algebra for the CP($n)$ Kazama-Suzuki models at
negative central charges. For the first three examples we checked this
explicitly on the level of structure constants. In the case $n>3$ at
least the self-coupling constants of the field of dimension two coincide
using the prediction mentioned above. Finally, we pointed out that this
algebra does not admit any unitary representation.
\meno
We would like to end with several open questions which quite naturally arise
given our results:
\item{(i)} Do other ${\cal SW}(1,2,...,n)$ algebras also admit solutions
 different from the super ${\cal W}_{n+1}$-algebra? (cf.\ ref.\ [13] in the
 $N=0$ case)
\item{(ii)} What are the unifying super ${\cal W}$-algebras for other
 Kazama-Suzuki models?
\item{(iii)} Are there unifying super ${\cal W}$-algebras admitting unitary
 representations?
\item{(iv)} Do Kazama-Suzuki models exhaust the classification of
 $N=2$ unitary models?
\item{(v)} Are there non-unitary minimal $N=2$ models?
\meno
{\bf Acknowledgements:}
It is a pleasure to thank
W.\ Eholzer, M.\ Flohr, R.\ H\"ubel,
N.\ Mohammedi, W.\ Nahm, M.\ R\"osgen, R.\ Schimmrigk
and R.\ Varnhagen for discussion,
M.\ Terhoeven for carefully reading the
manuscript, A.\ Honecker for making available his package {\it 'commute'}
and the MPI f\"ur Mathematik in Bonn for access to their computers.
\meno
\chapter{Appendix}
\meno
${\cal SW}(1,2,3)^{[{\rm I}]}$:
\smno
$\bullet$ the spin-$3$ generator of ${\cal W}(2,3)$ for
  $\tilde c={4(21-c)c\over c+27}$:
\smno
${\ts
{\sqrt {2(33-c)\over 5c+27}}{6c(c-21)
 \over(c-1)(c+6)(c+27)(2c-3)}
{\Big(}
{2(c-1)(c+6)(c+9)(2c-3)(5c-9)C_{\cal WW}^{\cal W}
 \over3c(c+3)(5c-12)(7c-27)}N({\cal W}_1J)
}$
\pano
${\ts
 +{(c+6)(c+9)(2c-3)(5c-9)C_{\cal VW}^{\cal W}\alpha_{\cal VWW}
\over18c(c-33)}{\cal V}_1
 -{(c-1)(c+6)(c+9)(2c-3)(5c-9)C_{\cal WW}^{\cal W}
    \over9(c+3)(5c-12)(7c-27)}
{\cal W}_4
}$
\pano
${\ts
+{c\over18}\pt^2J
+{c\over6}\pt L
-{c\over6}N(G^-G^+)
+N(LJ)
-{1\over c}N(N(JJ)J){\Big)}
}$
\hskip 114pt (A.1)
\smno
$\bullet$ the spin-$3$ generator of the CP($3$) part for
   $\tilde c={2(c+9)(5c-9)\over c+27}$:
\smno
${\ts
{\sqrt {6(3-c)(c+18)\over c(c-27)}}{3(c+9)(5c-9)\over(c-1)(c+6)
	   (c+27)(2c-3)}
{\Big(}
 {4(c-21)(c-1)(c+6)(2c-3)C_{\cal WW}^{\cal W}\over3(c+3)(5c-12)(7c-27)}
  N({\cal W}_1J)
}$
\pano
${\ts
+{c(c-21)(c+6)(2c-3)C_{\cal VW}^{\cal W}\alpha_{\cal VWW}
\over54(c-3)(c+18)}
{\cal V}_1
 -{2c(c-21)(c-1)(c+6)(2c-3)C_{\cal WW}^{\cal W}\over9(c+3)(5c-12)(7c-27)}
{\cal W}_4
}$
\pano
${\ts
 +{c\over18}\pt^2J
+{c\over6}\pt L
-{c\over6}N(G^-G^+)
+N(LJ)
-{1\over c}N(N(JJ)J){\Big)}
}$
\hskip 114pt (A.2)
\smno
$\bullet$ the spin-$4$ generator of the CP($3$) part for
   $\tilde c={2(c+9)(5c-9)\over c+27}$:
\smno
${\ts
 {\sqrt{54(27-7c)(c-21)(c+3)(c+9)(5c-9)\over
  (c-81)(c-27)(c-3)(c-1)(c+18)(c+27)(5c+27)}}
{\Big(}
{c(c+9)(1458-909c+84c^2+7c^3)
 C_{\cal WW}^{\cal W}\over72(3-2c){(c+3)}^2(c+6){(7c-27)}^2}\pt^3J
}$
\pano
${\ts
 +{-1944+1557c+373c^2\over c(c-1)(5c-12)
   (1377+246c+25c^2)}N(N({\cal W}_1J)J)
-{2(2430-4059c+1350c^2+223c^3)\over3(c+3)(c-1)
 (5c-12)(1377+246c+25c^2)}N({\cal W}_1L)
}$
\pano
${\ts
  +{c(c+9)(1458-909c+84c^2+7c^3)
 C_{\cal WW}^{\cal W}\over12{(c+3)}^2(c+6)(2c-3){(7c-27)}^2
 }N(\pt G^-G^+)
 -{(c-21)(c+9)(5c-9)C_{\cal WW}^{\cal W}
  C_{\cal VW}^{\cal W}\alpha_{\cal VWW}\over
  6{(c+3)}^2{(7c-27)}^2}N({\cal V}_1J)
}$
\pano
${\ts
 +{(c+9)(1458-909c+84c^2+7c^3)
   C_{\cal WW}^{\cal W}\over2{(c+3)}^2(c+6)(2c-3){(7c-27)}^2
   }N(N(G^-J)G^+)
 +{(c+9)(1458-909c+84c^2+7c^3)
   C_{\cal WW}^{\cal W}\over2(3-2c){(c+3)}^2(c+6){(7c-27)}^2
   }N(\pt LJ)
}$
\pano
${\ts
+{(c-6)(c+9)(1458-909c+84c^2+7c^3)
 C_{\cal WW}^{\cal W}\over
 12(3-2c){(c+3)}^2(c+6){(7c-27)}^2}
 N(G^-\pt G^+)
+{c(c-21)(c+9)(5c-9)C_{\cal WW}^{\cal W}
 C_{\cal VW}^{\cal W}\alpha_{\cal VWW}\over
 54{(c+3)}^2{(7c-27)}^2}
 {\cal V}_4
}$
\pano
${\ts
 -{(c-21)(c+9)(5c-9)(-135+140c+3c^2)C_{\cal WW}^{\cal W}
  \over3{(c+3)}^2{(7c-27)}^2(1377+246c+25c^2)}N({\cal W}_1{\cal W}_1)
-{2916-2457c+96c^2+29c^3\over12(c+3)(5c-12)(1377+246c+25c^2)}
\pt^2{\cal W}_1
}$
\pano
${\ts
 -{(c+9)(-3542940+2589408c+1216053c^2+265545c^3-552258c^4+
    1194c^5+4937c^6+141c^7)C_{\cal WW}^{\cal W}\over4c(c-1){(c+3)}^2(c+6)
    (2c-3){(7c-27)}^2(1377+246c+25c^2)}N(N(JJ)L)
}$
\pano
${\ts
+{(c+9)(590490-1496637c+1241676c^2-352494c^3+13866c^4+859c^5)
 C_{\cal WW}^{\cal W}\over6(c-1)(c+3)(c+6)(2c-3){(7c-27)}^2(1377+246c+25c^2)}
 N(LL)
 -{2\over(c+3)(5c-12)}N({\cal W}_4J)
}$
\pano
${\ts
 +{3(c+9)(-1062882+1200663c-59049c^2-13257c^3-74907c^4+6846c^5+
    346c^6)C_{\cal WW}^{\cal W}\over2c^2(c-1){(c+3)}^2(c+6)(2c-3){(7c-27)}^2
     (1377+246c+25c^2)}N(N(JJ)N(JJ))
}$
\pano
${\ts
+{3(c+9)(-1653372+685260c+122445c^2-1422c^3-17088c^4+770c^5+
     47c^6)C_{\cal WW}^{\cal W}\over4c{(c+3)}^2(c+6)(2c-3){(7c-27)}^2
     (1377+246c+25c^2)}N(N(LJ)J)
}$
\pano
${\ts
 -{(c+9)(2125764-3254256c+1198719c^2+108432c^3-87174c^4+
   1848c^5+107c^6)C_{\cal WW}^{\cal W}\over8c{(c+3)}^2(c+6)(2c-3){(7c-27)}^2
   (1377+246c+25c^2)
   }N(\pt J\pt J)
}$
\pano
${\ts
+{(c+9)(-1062882+2630961c-1045872c^2-89964c^3+47376c^4+
 987c^5+34c^6)C_{\cal WW}^{\cal W}\over(1377+246c+25c^2)
 12(3- 2c){(c+3)}^2(c+6){(7c-27)}^2}
\pt^2L
}$
\pano
${\ts
-{c\over6(c+3)(5c-12)}N({\cal W}_3G^-)
+{c\over6(c+3)(5c-12)}N({\cal W}_2G^+)
{\Big)}
}$
\hskip 139pt (A.3)
\meno
${\cal SW}(1,2,3)^{[{\rm II}]}$:
\smno
$\bullet$ the pair $\Phi^\pm$ of generic
  null fields of dimension ${\textstyle{11\over2}}$:
\smno
${\ts
\Phi^\pm\sim
\pm{36+180c+293c^2-8c^3-4c^4\over8c(c-1)(c+1)(c+6)(2c-3)}
   {\cal N}_s({\cal N}_s({\cal N}_s({\cal N}_s({\cal LL}){\cal L}){\cal L})
   D^\mp{\cal L})
}$
\pano
${\ts
\pm{288-810c+3919c^2-2379c^3-52c^4\over84c(c-1)(c+1)(c+6)(2c-3)}
   {\cal N}_s({\cal N}_s({\cal N}_s({\cal L}[D^-,D^+]{\cal L}){\cal L})
   D^\mp{\cal L})
}$
\pano
${\ts
+{33-4c\over2(c-1)(2c-3)}{\cal N}_s({\cal N}_s({\cal N}_s({\cal L}D^\mp\pt
   {\cal L}){\cal L}){\cal L})
\pm{-108+185c+38c^2\over42c(1-c)(c+1)}
   {\cal N}_s({\cal N}_s({\cal L}\pt^2{\cal L})D^\mp{\cal L})
}$
\pano
${\ts
+{2(144-234c+41c^2)C_{\cal VW}^{\cal W}\alpha_{\cal VWW}\over7c(c-24)(c-1)}
  {\cal N}_s({\cal V}D^\mp\pt{\cal L})
\pm{(c+1)(5c-12)C_{\cal WW}^{\cal W}C_{\cal VW}^{\cal W}\alpha_{\cal VWW}
   \over{c}(24-c)(c+3)}
   {\cal N}_s({\cal V}D^\mp{\cal W})
}$
\pano
${\ts
\pm{12(5c-12)C_{\cal VW}^{\cal W}\alpha_{\cal VWW}\over5c(c-24)(c-1)}
  {\cal N}_s({\cal N}_s({\cal V}D^\mp{\cal L}){\cal L})
\pm{18(4c+3)C_{\cal WW}^{\cal W}\over5c(3-2c)(c+3)(c+6)}
  {\cal N}_s({\cal N}_s({\cal N}_s({\cal W}D^\mp{\cal L}){\cal L}){\cal L})
}$
\pano
${\ts
+{3(-648-1440c+939c^2+71c^3)C_{\cal WW}^{\cal W}\over
   5c(3-2c)(c+3)(c+6)(5c-12)}
   {\cal N}_s({\cal N}_s({\cal W}\pt{\cal L})D^\mp{\cal L})
\pm{1\over{c}}
   {\cal N}_s({\cal N}_s({\cal WW})D^\mp{\cal L})
}$
\pano
${\ts
\pm{(3672+3852c-4254c^2+114c^3+95c^4)C_{\cal WW}^{\cal W}\over
   5c(c+3)(c+6)(2c-3)(5c-12)}
   {\cal N}_s({\cal W}D^\mp\pt^2{\cal L})
+{\cal N}_s({\cal W}D^\mp\pt{\cal W})
}$
\pano
${\ts
\pm{7(-216+144c-93c^2+53c^3)C_{\cal WW}^{\cal W}\over5c(c+3)(c+6)(2c-3)
    (5c-12)}
   {\cal N}_s({\cal N}_s({\cal W}[D^-,D^+]{\cal L})D^\mp{\cal L})
}$
\pano
${\ts
+{48(-324+81c+138c^2+7c^3)C_{\cal WW}^{\cal W}\over25c(c+3)(c+6)(2c-3)
  (5c-12)}
   {\cal N}_s({\cal N}_s({\cal W}D^\mp\pt{\cal L}){\cal L})
}$
\hskip 155pt (A.4)
\smno
$\bullet$ the spin-$3$ generator of ${\cal W}(2,3)$ for $\tilde c=-2$:
\smno
${\ts
 {\sqrt {6(c-24)\over c}}{6\over(c-1)(c+6)(2c-3)}
{\Big(}
 {2(c-1)(c+1)(c+6)(2c-3)C_{\cal WW}^{\cal W}\over3c(c+3)(5c-12)}
  N({\cal W}_1 J)
}$
\pano
${\ts
 +{(c+1)(3-2c)(c+6)C_{\cal VW}^{\cal W}\alpha_{\cal VWW}\over6(c-24)}
 {\cal V}_1
 +{(1-c)(c+1)(c+6)(2c-3)C_{\cal WW}^{\cal W}\over9(c+3)(5c-12)}
{\cal W}_4
}$
\pano
${\ts
 +{c\over18}\pt^2J+{c\over6}\pt L-{c\over6}N(G^-G^+)
+N(LJ)
-{1\over c}N(N(JJ)J){\Big)}
}$
\hskip 114pt (A.5)
\meno
\chapter{References}
\meno
\baselineskip=11.5pt
\bibitem{1} C.\ Ahn,
{\it Free superfield realization of $N=2$ quantum super $\ca W_3$-algebra},
Mod.\ Phys.\ Lett.\ {\bf A9} (1994) 271

\bibitem{2} D.\ Altschuler,
{\it Quantum equivalence of coset space models},
Nucl.\ Phys.\ {\bf B305} (1988) 685

\bibitem{3} I.\ Bars, {\it Heterotic superstring vacua in four dimensions
based
on non-compact affine current algebras}, Nucl.\ Phys.\ {\bf B334} (1990) 125

\bibitem{4} M.\ Bershadsky, W.\ Lerche, D.\ Nemeschansky, N.P.\ Warner,
{\it Extended $N=2$ superconformal structure of gravity and
${\cal W}$-gravity coupled to matter}, Nucl.\ Phys.\ {\bf B401} (1993) 304

\bibitem{5} R.\ Blumenhagen, M.\ Flohr, A.\ Kliem, W.\ Nahm, A.\ Recknagel,
R.\ Varnhagen, {\it ${\cal W}$-algebras with two and three generators},
Nucl.\ Phys.\ {\bf B361} (1991) 255

\bibitem{6} R.\ Blumenhagen, {\it $N=2$ supersymmetric \w-algebras},
Nucl.\ Phys.\ {\bf B405} (1993) 744

\bibitem{7} R.\ Blumenhagen, W.\ Eholzer, A.\ Honecker, K.\ Hornfeck,
R.\ H\"ubel,
{\it Cosets and unifying \w-algebras},
Phys.\ Lett.\ {\bf 332B} (1994) 51

\bibitem{8} R.\ Blumenhagen, W.\ Eholzer, A.\ Honecker, K.\ Hornfeck,
R.\ H\"ubel,
{\it Coset realization of unifying \w-algebras},
preprint BONN-TH-94-11 (hep-th/9406203)

\bibitem{9} P.\ Bowcock, P.\ Goddard,
{\it Coset constructions and extended conformal algebras},
Nucl.\ Phys.\ {\bf B313} (1989) 293

\bibitem{10}
P.\ Fendley, W.\ Lerche, S.D.\ Mathur, N.P.\ Warner,
{\it $N=2$ supersymmetric integrable models from affine Toda theories},
Nucl.\ Phys.\ {\bf B348} (1991) 66

\bibitem{11} D.\ Gepner, {\it Space-time supersymmetry in compactified string
theory and superconformal models}, Nucl.\ Phys.\ {\bf B296} (1988) 757

\bibitem{12} A.\ Honecker, {\it A note on the algebraic evaluation of
correlators in local chiral conformal field theories}, preprint
BONN-HE-92-25 (hep-th/9209029)

\bibitem{13} K.\ Hornfeck, {\it ${\cal W}$-algebras with a set of primary
fields of dimension $(3,4,5)$ and $(3,4,5,$ $6)$},
Nucl.\ Phys.\ {\bf B407} (1993) 237

\bibitem{14} T.\ Inami, Y.\ Matsuo, I.\ Yamanaka,
{\it Extended conformal algebra with $N=2$ supersymmetry},
Int.\ Jour.\ Mod.\ Phys.\ {\bf A5} (1990) 4441

\bibitem{15} K.\ Ito, {\it Quantum hamiltonian reduction
and $N=2$ coset models}, Phys.\ Lett.\ {\bf 259B} (1991) 73

\bibitem{16} K.\ Ito, {\it $N=2$ superconformal CP(n) model},
Nucl.\ Phys.\ {\bf B370} (1992) 123

\bibitem{17} K.\ Ito, {\it Free field realization of $N=2$ super
${\cal W}_3$-algebra}, Phys.\ Lett.\ {\bf B304} (1993) 271

\bibitem{18} K.\ Ito, H.\ Kanno, {\it Lie superalgebra and extended
topological conformal symmetry in non-critical ${\cal W}_3$-strings},
UTHEP-277, May 1994 (hep-th/9405049)

\bibitem{19} E.\ Ivanov, S.\ Krivonos,
{\it Superfield realization of $N=2$ super-${\cal W}_3$},
Phys.\ Lett.\ {\bf 291B} (1992) 63

\bibitem{20} Y.\ Kazama, H.\ Suzuki, {\it New $N=2$
superconformal field theories and superstring compactification},
Nucl.\ Phys.\ {\bf B321} (1989) 232

\bibitem{21} Y.\ Kazama, H.\ Suzuki,
{\it Bosonic construction of conformal field theories with extended
supersymmetry},
Mod.\ Phys.\ Lett.\ {\bf A4} (1989) 235

\bibitem{22} H.\ Lu, C.N.\ Pope, L.J.\ Romans, X.\ Shen, X.J.\ Wang,
{\it Polyakov construction of the $N=2$ super $\ca W_3$-algebra},
Phys.\ Lett.\ {\bf 264B} (1991) 91

\bibitem{23} D.\ Nemeschansky, S.\ Yankielowicz,
{\it $N=2$ ${\cal W}$-algebras, Kazama-Suzuki models and Drinfeld-Sokolov
reduction}, preprint USC-91/005

\bibitem{24} R.E.C.\ Perret, {\it Dual formulation of classical
${\cal W}$-algebras}, Lett.\ Math.\ Phys.\ 27 (1993) 27

\bibitem{25} R.E.C.\ Perret,
{\it A classical $N=4$ super ${\cal W}$-algebra},
Int.\ J.\ Mod.\ Phys.\ {\bf A8} (1993) 3615

\bibitem{26} L.J.\ Romans, {\it The $N=2$ super ${\cal W}_3$-algebra},
Nucl.\ Phys.\ {\bf B369} (1992) 403

\bibitem{27} P.\ West, {\it 'Introduction to supersymmetry
and supergravity'}, second edition 1990,
${}$ World Scientific, Singapore

\bibitem{28} A.\ Wi{\ss}kirchen, {\it Konstruktion von $N=2$
Super-${\cal W}$-Algebren}, Diplomarbeit BONN 1994

\bibitem{29} S.\ Wolfram, {\it 'Mathematica'}, Addison-Wesley, Reading, MA
(1991)

\bibitem{30} A.B.\ Zamolodchikov, {\it Infinite additional symmetries in two
dimensional conformal quantum field theory}, Theor.\ Math.\ Phys.\
65 (1986) 1205

\vfill
\end